\documentclass{IEEEtran4PSCC}

\usepackage{cite}
\usepackage{amsmath}

\usepackage[pdftex]{graphicx}
\usepackage{svg}
\usepackage{float}

\usepackage{amsmath}
\usepackage{subcaption}
\usepackage{float}
\makeatletter
\let\old@ps@headings\ps@headings
\let\old@ps@IEEEtitlepagestyle\ps@IEEEtitlepagestyle
\def\psccfooter#1{%
    \def\ps@headings{%
        \old@ps@headings%
        \def\@oddfoot{\strut\hfill#1\hfill\strut}%
        \def\@evenfoot{\strut\hfill#1\hfill\strut}%
    }%
    \def\ps@IEEEtitlepagestyle{%
        \old@ps@IEEEtitlepagestyle%
        \def\@oddfoot{\strut\hfill#1\hfill\strut}%
        \def\@evenfoot{\strut\hfill#1\hfill\strut}%
    }%
    \ps@headings%
}
\makeatother

\psccfooter{%
        \parbox{\textwidth}{\hrulefill \\ \small{22nd Power Systems Computation Conference} \hfill \begin{minipage}{0.2\textwidth}\centering \vspace*{4pt} \includegraphics[scale=0.06]{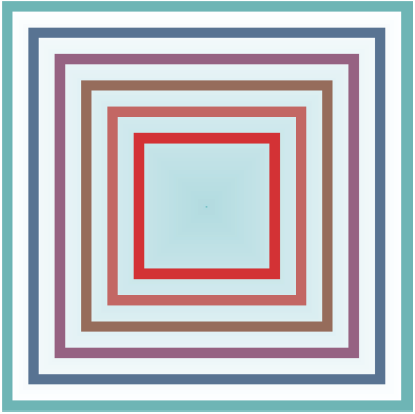}\\\small{PSCC 2022} \end{minipage} \hfill \small{Porto, Portugal --- June 27 -- July 1, 2022}}%
}

\usepackage{amsthm}
\usepackage{amssymb}
\usepackage{booktabs}
\usepackage{url}
\usepackage{siunitx}
\usepackage{hyperref}
\usepackage{enumitem}
\usepackage{algorithm2e}
\newcommand{\kaist}{\textsc{Kaist}-agent}
\newcommand{\parl}{\textsc{PARL}-agent}
\newcommand{\dqn}{\textsc{D3QN}-agent}
\newcommand{\nanyang}{\textsc{Nanyang}-agent}
\RestyleAlgo{ruled}
\SetKwInput{KwInput}{Input}
\SetKwInput{KwReturn}{Return}
\SetKwInput{KwInit}{Initialize}
\SetKw{Continue}{continue}

\begin{document}
\title{Improving Robustness of Reinforcement Learning for Power System Control with Adversarial Training}

\author{Alexander Pan$^{1}$, Yongkyun (Daniel) Lee$^{1}$, Huan Zhang$^{2}$, Yize Chen$^{3}$ and Yuanyuan Shi$^{4}$
\thanks{$^{1}$Alexander Pan and Yongkyun (Daniel) Lee are with the Computing and Mathematical Sciences, California Institute of Technology, {\tt\small aypan.17@gmail.com}.}%
\thanks{$^{2}$Huan Zhang is with the Department of Computer Science, CMU.}%
\thanks{$^{3}$Yize Chen is with the Computational Research Division, LBL, Berkeley.}%
\thanks{$^{4}$Yuanyuan Shi is with the Department of Electrical and Computer Engineering, University of California San Diego.}%
}

\maketitle

\begin{abstract}
Due to the proliferation of renewable energy and its intrinsic intermittency and stochasticity, current power systems face severe operational challenges. Data-driven decision-making algorithms from reinforcement learning (RL) offer a solution towards efficiently operating a clean energy system. Although RL algorithms achieve promising performance compared to model-based control models, there has been limited investigation of RL robustness in safety-critical physical systems. In this work, we first show that several competition-winning, state-of-the-art RL agents proposed for power system control are \emph{vulnerable} to adversarial attacks. 
Specifically, we use an adversary Markov Decision Process to learn an attack policy, and demonstrate the potency of our attack by successfully attacking multiple winning agents from the Learning To Run a Power Network (L2RPN) challenge~\cite{yoon2021winning,liu2020parl}, under both white-box and black-box attack settings.
We then propose to use \emph{adversarial training} to increase the robustness of RL agent against attacks and avoid infeasible operational decisions.
To the best of our knowledge, our work is the first to highlight the fragility of grid control RL algorithms, and contribute an effective defense scheme towards improving their robustness and security.
\end{abstract}


\thanksto{\noindent
1. Alexander Pan and Yongkyun (Daniel) Lee are with the Computing and Mathematical Sciences, Caltech. Corresponding author: {\tt\small aypan.17@gmail.com}.\\2. Huan Zhang is with the Department of Computer Science, CMU.\\3. Yize Chen is with the Computational Research Division, LBNL, Berkeley.\\4. Yuanyuan Shi is with the Department of Electrical and Computer Engineering, University of California San Diego.}

\section{Introduction}
\label{sec:intro}
As power systems shift towards renewable energy sources, system operators are facing emerging operational challenges. Human operators may find it challenging to manage the intrinsic stochasticity of renewable energy generation, motivating research for autonomous power systems control schemes. In addition to such operation challenges, for active distribution networks, designing a model-based control algorithm usually requires extensive domain knowledge and great amount of work on identifying the grid topology and network parameters~\cite{deka2016estimating}, while model misspecification can cause significant impacts on system operation. These difficulties spurred the development of model-free data-driven control pipelines, specifically deep reinforcement learning (RL) algorithms~\cite{mnih2013playing, mnih2016asynchronous, marot2020whitepaper}. As a result, there are several works that apply RL algorithms to various of control and operation tasks in power systems \cite{kelly2020pn}, ranging from topology control~\cite{grid2op,yoon2021winning,liu2020parl}, PV control~\cite{cao_multi-agent_2020}, and Volt-VAR control~\cite{liu2020two, wang_safe_2020,duan_deep-reinforcement-learning-based_2020}.   

\begin{figure}[!tb]
\centering
\includegraphics[scale=0.39]{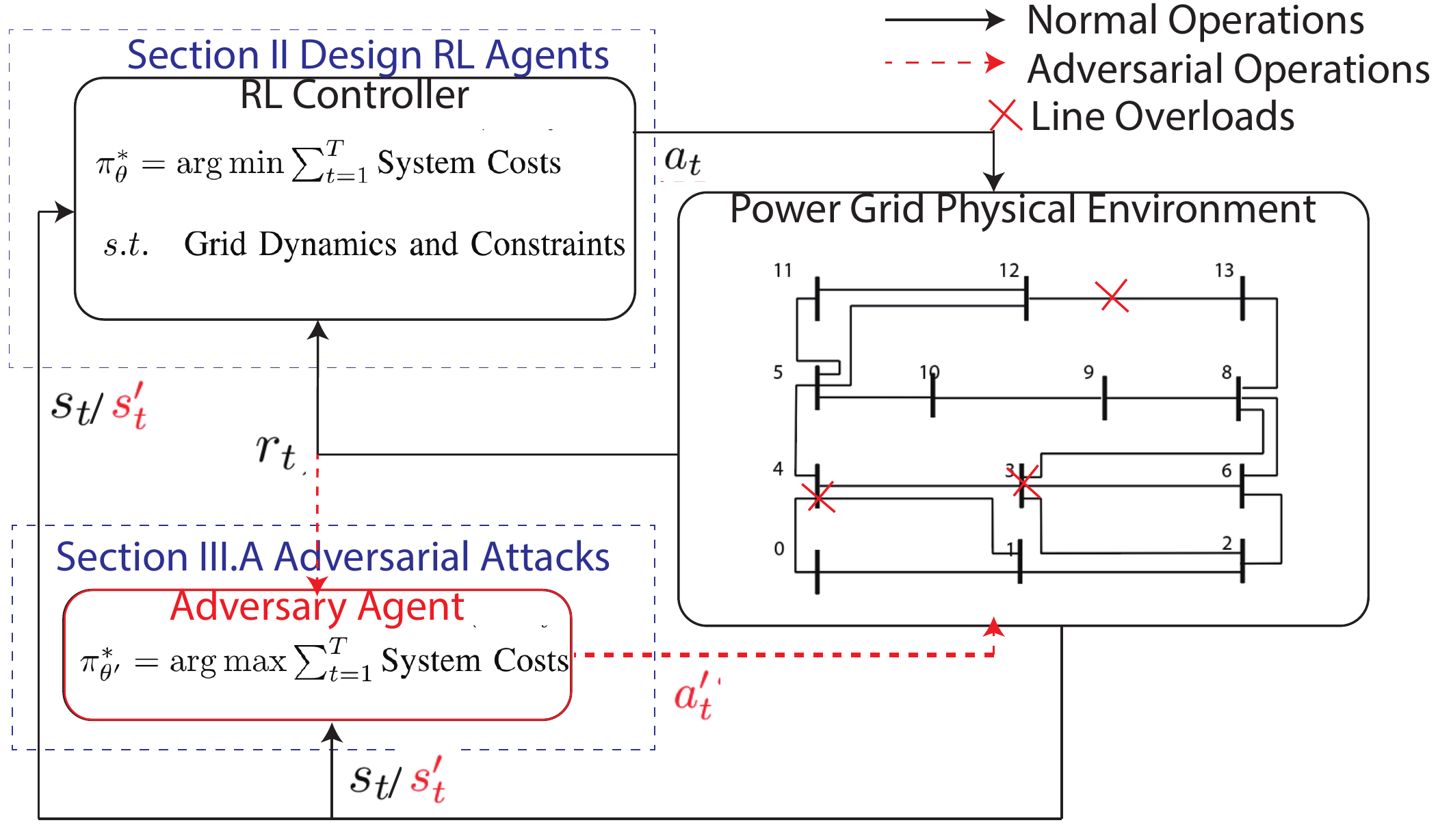}
\caption{We use RL to both accomplish the power system control task and find the adversarial action.  The adversary learns to attack the agent by disconnecting lines and observing their effect. We require only black-box access to the agent. Attack actions and states are denoted as $a_t'$ and $s_t'$ respectively. We also make use of the adversary agent for adversarial training to provide better robustness (Section \ref{sec:advtraining_algo}). } 
\label{fig:method}
\end{figure}

However, in order to deploy such data-driven agents to safety-critical power system tasks, there is an urgent need to investigate whether the learned agents can be \emph{robust} under a various operating scenarios, i.e., feasible power flow even when a power line is temporarily disconnected due to maintenance, environmental hazards, or cyber-physical attacks~\cite{rushe_borger_2021}. 
Previous works in machine learning have demonstrated that standard RL agents are vulnerable to observation-noise attacks. With small perturbations over state space or minor alterations on the underlying dynamics, even fully-optimized RL agents will output non-optimal actions~\cite{zhang2020robustdrl, chen2021attack}. 
Meanwhile, power grids have always been regarded as safety-critical infrastructures~\cite{usenergy2016}, and it is of top priority to validate the reliability of proposed algorithms under either system state uncertainties (e.g., renewable forecasting~\cite{douglas1998risk}) or system uncertainties (e.g., N-1 security criterion~\cite{vrakopoulou2013probabilistic, marot2020whitepaper}). Therefore, it is crucial for learning-based control schemes to be robust against exogenous events, whether or not they are malicious in origin. 



In this work, we focus on a typical power grid topology control task introduced by the Learning To Run a Power Network (L2RPN) challenge~\cite{yoon2021winning,liu2020parl} as a standard testbed for grid operation RL algorithm development. The designed controllers observe the underlying state of the power grid (network topology and parameters, load injections, and etc.), and take economical actions (modifying network topology, decide power setpoints of generators, etc.) to ensure feasible power flow. Robustness is difficult to achieve in this task due to the nonlinear, complex spatiotemporal dynamics within the power grids. In one of our test case on IEEE 118 bus system~\cite{christie2000power}, there are around $3.88 \cdot 10^{76}$ network topologies and $9.81 \cdot 10^{55}$ power line actions. When power lines are disconnected, RL agents struggle to adapt; in real life such a failure would lead to a blackout.

In this paper, we study the impacts of adversarial attacks on RL agent and develop an effective defense method to enhance its robustness. We affirmatively answer two questions: 

1) \emph{Is it possible to learn a harmful adversary?}  

2) \emph{Does adversarial training improve agents' robustness?}

At a high level, we first use RL to train an adversary to disconnect vital lines, by rewarding it for reducing the reward of grid operation RL agent. Then, we use our learned adversary, in turn, to increase the robustness of the grid operation RL agent, through adversarial training. 
The adversary generates training scenarios that are more difficult than normally encountered, thus boosting the robustness of the resulting RL algorithm. An overview of our approach is described in Figure~\ref{fig:method}.

We instantiate our RL agents in the Learning to Run a Power Network (L2RPN) environment~\cite{marot2020whitepaper}. The environment, similar to OpenAI Gym~\cite{openaigym}, simulates realisitic power networks and provides an interface to train RL agents to operate them. 
An agent deployed on the L2RPN environment manages the power network topology while subjected to maintenance and environmental hazards. These hazards temporarily decommission power lines and force the agent to modify the network to prevent blackouts. 
\emph{We make use of the winning algorithms from recent L2RPN competitions~\cite{liu2020parl, yoon2021winning, yan2020wcci} to show the vulnerability of normal RL agents, and show the improved robustness provided by adversarial training.}

Our contributions are summarized as follows.
\begin{itemize}[leftmargin=*]
    \item We propose an agent-specific adversary MDP to learn an adversarial policy for a given agent. We demonstrate the effectiveness of our adversaries by conducting both white-box and black-box transfer attacks against the \kaist, \parl, and \nanyang~(winning agents from previous L2RPN challenges), which lead to over 90\% performance drop of the winning agents, and greatly outperform other baselines attack methods. 
    \item We use our learned adversary to improve the robustness of RL algorithm via adversarial training. Adversarial training exposes the grid operation RL agent to more difficult scenarios than normally seen. As a result, when being attacked, it is able to quickly respond and re-balance the grid. We instantiate the proposed method in \kaist, and show significant improvement on both the clean (no adversary) and robustness performance (with different adversaries).
\end{itemize}


Our work is the first attempt to improve robustness for a general class of RL algorithms, and takes a step towards safe deployment of RL controllers for power grids. Compared to existing optimization-based robust power system operation methods, which require heavy modeling and specific solution techniques~\cite{bertsimas2012adaptive, roald2017chance}, our proposed framework directly learns a mapping from power system states to operation actions.

Our paper is organized as follows. Section II describes related work. Section III describes how to frame power systems operations as a Markov Decision Process (MDP). Section IV describes our proposed adversary MDP for learning a strong adversary and adversarial training to improve robustness. Section V describes our experimental results. Section VI provides a discussion and concluding remarks.


\section{Power Network Operation as a Markov Decision Process}\label{sec:mdp}
In this section, we first introduce the power system operation model considered in this paper. Then, we show how reinforcement learning can serve as an ideal framework for efficiently solving such operation problem. 

\subsection{Power System Operation Problem}
We consider a power system where $\mathcal{N}, \mathcal{L}$ denote the set of nodes and lines, and $|\mathcal{N}| = n, |\mathcal{L}| = l$. We define the set of nodes with generators as $\mathcal{G}$, and without loss of generality, we consider $|\mathcal{N}| = |\mathcal{G}| = n$. We denote the active/reactive power outputs of generator at node $i \in \mathcal{N}$ as $p_{G, i}, q_{G, i}$ (which is controllable by the system operator). If there is no physical generators at node $i$, we just simply restrict $p_{G, i} = q_{G, i} = 0$. Define the demand at node $i$ as $p_{D, i}, q_{D, i}$, which is given by the environment. We look into a power grid operational setting where topology can be reconfigured to minimize the system costs. Define the topology choice $\Omega \in \mathcal{S}_{\Omega}$ where $\mathcal{S}_{\Omega}$ is the set of all possible topologies by line switching, and define $\mathcal{L}_{\Omega}$ as all lines connected under topology $\Omega$. $G_{\Omega}$ and $B_{\Omega}$ are the conductance and susceptance matrices associated with topology $\Omega$.
If line $l_{ij} \notin \mathcal{L}_{\Omega}$, $G_{\Omega, i, j} = B_{\Omega, i, j} = 0$.

We follow the power grid operational cost definition in the L2RPN challenge~\cite{marot2020whitepaper}, where the total system cost over horizon $T$ is defined as 
\begin{equation}\label{eq:opt_obj}
    c(e) = \sum_{t=1}^{T} c_{\text{operations}} (t) 
\end{equation}
and $c_{operations}(t)$ tracks the system operation cost at each time step,
\[c_\text{operations}(t) = \mathcal{E}_\text{loss}(t) + \alpha \cdot  \mathcal{E}_\text{redispatch}(t),\]
where $\mathcal{E}_\text{loss}(t)$ is the sum of the energy loss (due to transmission line resistance) and \[\mathcal{E}_\text{redispatch}(t) \propto \sum_{i \in \mathcal{N}} |p_{G, i}(t)-p_{G, i}(t-1)|\] is the redispatching cost. Here $\alpha$ is a parameter trading off the two operational cost terms, and we set $\alpha = 1$ in experiments.

The goal of the system operator is to reduce the total system operation costs, subject to system dynamics represented by the power flow equations, 
\begin{subequations}
\label{eqn:main}
\begin{align}
\min_{p_G, q_G, \Omega} \quad & c(e) \\
\text { s.t. }  \quad &  G(t+1), B(t+1) = f(G(t), B(t), \Omega(t))\,, \forall t \label{eq:rl_topology}\\
& g(\theta(t), v(t), p(t), q(t); G(t), B(t)) = 0\,,  \forall t  \label{eq:rl_pf}\\
& \theta(t) \in \mathcal{\theta}, v(t) \in \mathcal{V}\,, \forall t \label{eq:state_constr1}\\
& p_G(t) \in \mathcal{P}_G, q_G(t) \in \mathcal{Q}_G\,, \forall t \label{eq:gen_constr}\\
& p_{ij}(t) \in \mathcal{P}, q_{ij}(t) \in \mathcal{Q}\,, \forall t \label{eq:state_constr2} 
\end{align}
\end{subequations}
where~\eqref{eq:rl_topology} represents how the system dynamics change due to topology control action $\Omega(t)$. \eqref{eq:rl_pf} is the concise representation of the power flow equations \cite{roald2017chance}, in which $\theta$ is the voltage angle and $v$ is the voltage magnitude across all nodes. $p$ is the real power injection vector where $p_i = p_{G, i} -p_{D, i}$, and $q$ is the reactive power injection vector where $q_i = q_{G, i} -q_{D, i}$. Finally, $p_{ij}$ and $q_{ij}$ are the real and reactive power flow on branch $\{i, j\}$.
\eqref{eq:state_constr1}, \eqref{eq:state_constr2} and \eqref{eq:gen_constr} summarize the power system operation constraints on voltage angle, magnitude, generator capacity and line flow, respectively. Note that the optimization variables include both the topology choice $\Omega(t)$ and power dispatch $p_G(t), q_G(t)$ for all time steps $t = 1, ..., T$.

The optimal topology control and generation dispatch problem defined in \eqref{eqn:main} is a nonlinear, mixed-integer optimization problem, which is challenging to solve via conventional optimization techniques. Actually, it includes both continuous optimization variable $p_G, q_G$ and discrete optimization variable $\Omega(t)$ (topology choice), which leads to further complexity. In addition, even if sub-optimal solutions are acceptable, one needs the exact system model to solve the optimization, e.g., $G_{\Omega(t)}, B_{\Omega(t)}$, which are often unknown or hard to estimate in real systems~\cite{chen2020data}. 





\subsection{Reinforcement Learning for Power System Operation}
Reinforcement learning (RL) provides a powerful paradigm for solving \eqref{eqn:main}, by training a policy that maps the states to actions, so as to minimize the loss function defined as~\eqref{eq:opt_obj}. 
For the remaining of this section, we outline how the power network operation problem defined in \eqref{eqn:main} can be modeled as a RL problem. First, we define a Markov Decision Process (MDP) of 4-tuple $(\mathcal{S}, \mathcal{A}, \mathcal{P}, r)$ to represent the power network operation model. $\mathcal{S}$ is the set of states (which include network topology, generation and load values, and line flows), $\mathcal{A}$ is the set of agent actions (described below), $\mathcal{P} : \mathcal{S} \times \mathcal{A} \times \mathcal{S} \rightarrow \mathbb{R}$ is the transition probability (determined by the power flow equations), and $r$ is the reward function (described below). 
\begin{figure}[t]
     \centering
     \includegraphics[scale=0.425]{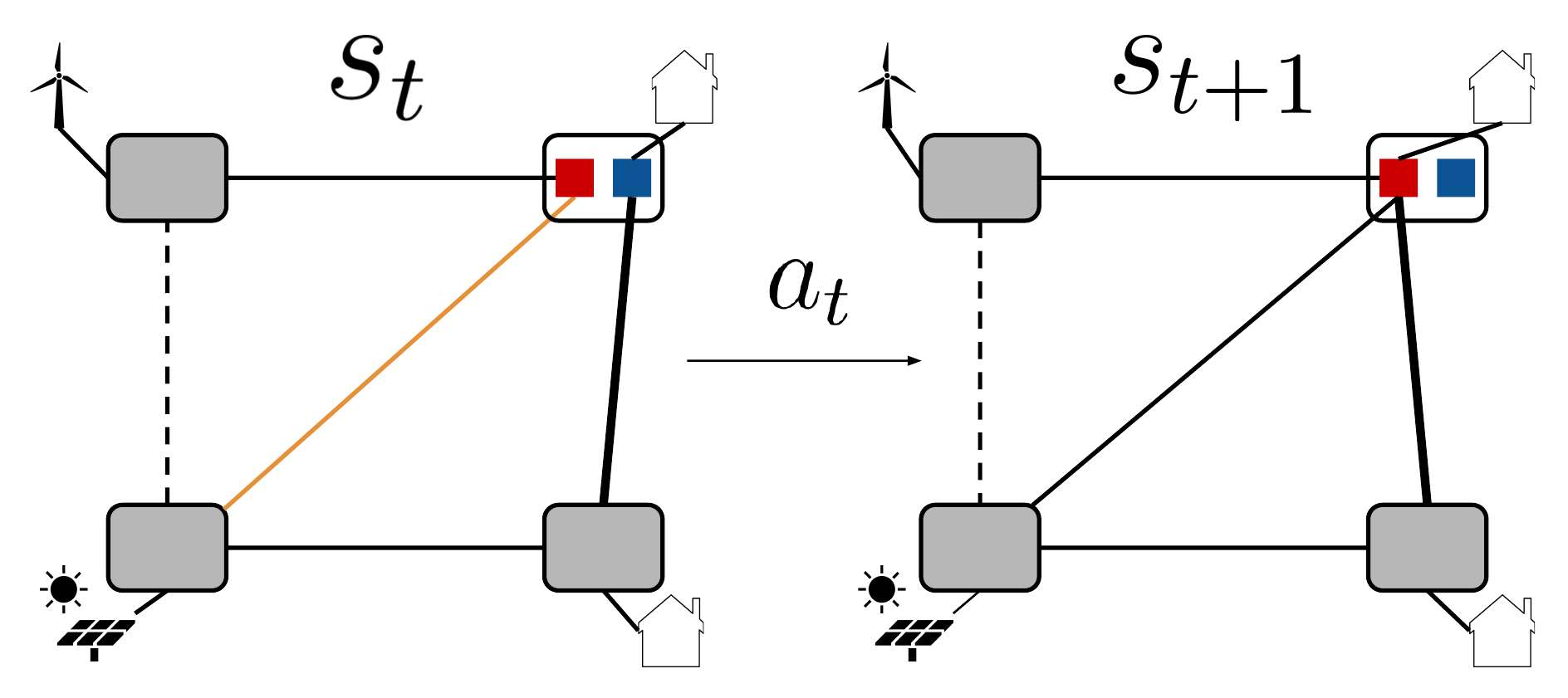}
    \caption{An example of a power network and a topology change. Power lines connect substations (rectangles), which either have loads (houses) or generators (wind or solar) attached. In state $s_t$, the dashed line is disconnected, causing the orange line to carry too much power. Action $a_t$ changes the network topology, connecting the bold line from the blue bus to the red bus. The grid then evolves to state $s_{t+1}$, where the (formerly) orange line's power is within its thermal limits.}
    \label{fig:bus}
\end{figure}

\noindent\textbf{Agent Action Space.} 
For our RL agents,  action $a = (a_{\Omega}, a_{\mathcal{G}})$ consists of two parts:
\begin{enumerate}
    \item \textit{Topological changes $a_{\Omega}$.} These actions alter the network topology $\Omega$ by changing the connections between power lines. Each substation has two buses, which can connect incoming power lines. A power line can be attached to either bus 1, bus 2, or neither bus. By switching lines on and off bus 1 and bus 2, an operator can effectively control the topology of the network (see Figure~\ref{fig:bus}). These are discrete actions.
    \item \textit{Redispatching $a_{\mathcal{G}}$.} An operator can also modify the generators' power setpoints $p_{G, i}, q_{G, i} \forall i \in \mathcal{N}$ (subject to physical constraints) across the network. These are continuous actions.
\end{enumerate}

\noindent\textbf{Reward function.} 
At each timestep, reward function is defined following the L2RPN environment~\cite{marot2020whitepaper}, 
\begin{equation}\label{eq:rl_reward}
r(s_t, a_t) = \begin{cases}
C-c_{\text{operations}} (t) & \text{normal operation},\\
0 & \text{blackout}.
\end{cases}
\end{equation}
The reward function is designed in a way such that, if the grid is under normal operation condition, $C-c_{\text{operations}} (t) > 0$, and higher reward will be given to encourage feasible, more economic power dispatch. If blackout happens, it will incur a very high operation cost (i.e., proportional to the total network load that failed to be served). In particular, we define $C=c_{\text{operations}}$ during a blackout,  obtaining the reward function form in \eqref{eq:rl_reward}. To summarize, the RL for optimal power network operation problem is as follows,
\begin{subequations}
\label{eqn:rl_operation}
\begin{align}
\max_{\theta} \quad & J(\theta):= E_{\mathcal{P}, \pi} \left[\sum_{t=1}^{T} r(s_t, a_t)\right] \\
\text{s.t.} \quad & a_t = \pi_{\theta}(s_t),\\
& a \in \mathcal{A}, s \in \mathcal{S}.
\end{align}
\end{subequations}
To solve Eq~\ref{eqn:rl_operation}, typical RL algorithms either estimate the expected reward of a state-action pair (value-based methods) or directly update the policy to minimize the reward (policy gradient)~\cite{sutton2018rl}. Yet directly estimating the value is difficult to scale up to a large number of states and actions, due to the combinatorial nature of state-action pairs. Indeed, we tried D3QN~\cite{wang2016d3qn}, a value-based method, and PPO~\cite{schulman2017ppo}, a policy gradient algorithm, and found that PPO to be a better optimizer. Thus, we use policy gradient estimation to solve our optimization. 

We also note that current RL agents considered for power system operations only consider clean measurements and safe operating conditions. In the next section, we will show such trained agents are vulnerable to adversaries, while the compromised policy can easily output actions that can cause infeasible power flow or even blackouts.

\section{Proposed Attacks and Defense}
In this section, we first formulate the attack problem as an adversary MDP, and propose two training methods under different information settings. Then we discuss how to use the learnt adversaries to enhance the robustness of grid operation RL agent via adversarial training.
\vspace{-3pt}
\subsection{Learning to attack}
\label{sec:adv_mdp}
Given an agent policy $\theta$, we wish to learn an adversarial policy $\theta'$ such that the normally trained policy will output ``unsafe'' actions. We do so by  solving the \emph{adversary MDP} $(\mathcal{S}, \mathcal{A}', \mathcal{P}_{\theta}, r')$.
Similar as the grid operating agent MDP, $\mathcal{S}$ is the set of all network states, which include network topology, generation and load values, and line flows. 
$\mathcal{A}'$ is the set of all available adversary actions. Specifically, the adversary's action space consists of the set of available power lines $\mathcal{L}_{adv} \subset \mathcal{L}$ to disconnect. In addition, we restrict that the adversary can only attack once every $k$ steps, to avoid incessant blackouts during training as well as reflect practical constraints on adversary's ability. $\mathcal{P}_\theta : \mathcal{S} \times \mathcal{A}' \times \mathcal{S} \rightarrow \mathbb{R}$ is the transition probability under grid control policy $\pi_{\theta}$ and adversary policy $\pi_{\theta'}$, and $r'$ is the adversary reward function, which is the negative of the operation agent's reward. The adversary seeks to minimize the expected reward of the grid operating agent, 
\begin{subequations}
\label{eqn:adv_rl}
\begin{align}
\min_{\theta'} \quad & J(\theta, \theta') \\
\text { s.t. } \quad & s_{t+1} \sim \mathcal{P}(s_{t+1}|s_t, a_t, a_t')\\
& a_t' = \pi_{\theta'}(s_t)\,, a_t = \pi_{\theta}(s_t) \text{ fixed $\theta$}\\
& a' \in \mathcal{A}', a \in \mathcal{A}, s \in \mathcal{S}
\end{align}
\end{subequations}
We consider two attack setups: \textit{white-box attacks} and \textit{black-box attacks}. Both assume attack agent has access to interacting with the power system environment, but only white-box attacks assume access to the agent's policy parameters.

\noindent\textbf{White-box attacks.} 
In this setup, we consider the attack agent has access to the grid operation RL agent's policy parameters. While this setting might not be realistic in practice, it represents an upper bound of the strength of learned adversaries.

\noindent\textbf{Black-box transfer attacks.}
In this setup, we train our own copy of grid operation RL agent and then proceed as in the white-box attack. Because the adversary does not know the exact agent's policy, it is not as strong as a white-box attacker. However, we found that as long as the adversary is trained against a strong agent, it can become a strong adversary and is able to \emph{transfer} its attacking ability across different agents. We demonstrate our adversary's transfer performance by training on one type of agent and attacking another type of agent (agents trained with different RL algorithms). This shows that our adversaries are not learning pathological behavior specific to a single agent, and are instead learning strong attacks with malicious physical consequences for power grids.

\subsection{Learning to defend}\label{sec:advtraining_algo}
 Given that these adversarial attacks are real threats to power grid operations, a natural goal is to train the RL agent to be as robust and reliable as possible subject to malicious behaviors. Essentially, we want our data-driven agents to interact with such adversarial scenarios during the training process, so that the resulting agents are robust against possible attacks. Mathematically, the robust RL training problem is defined as,
\begin{equation}\label{eqn:minmax}
   \max_{\theta} \min_{\theta'} J(\theta, \theta')\,,
\end{equation}
subject to the constraints given in Equation~\eqref{eqn:adv_rl}. To solve the robust reinforcement learning problem Eq~\eqref{eqn:minmax}, one can iteratively training the adversary policy $\theta'$ and the agent policy $\theta$. This method is known as robust adversarial reinforcement learning (RARL)~\cite{pinto2017robust}. However, there has been empirical evidence from deep reinforcement learning literature~\cite{gleave2020Adversarial} showing that iteratively training the agent and adversary converges slowly and does not confer greater robustness. 

An alternative way is to fix an adversary policy $\theta'$, and learn an agent policy $\theta$ that maximizes the expected system cost under the given adversary, 
\begin{equation}
    \max_{\theta} J(\theta, \theta'),
\end{equation}
We demonstrate that with a fixed adversary to perturb the environment during training, the robustness of RL agent can be greatly improved. We leave solving the full max-min problem, via iterative agent and adversary learning as a future work. 
Algorithm~\ref{alg:adv} describes our adversarial training procedure. 
\begin{algorithm}
\SetAlgoLined
\KwInput{Env $\mathcal{E}$; Trained adversary $\theta'$; Epochs $N$}
\KwInit{Grid operation agent parameters $\theta$;} 
 \For{$i = 1, \ldots, N$}{
    Store $\theta_i \leftarrow \theta_{i-1}$
    
    Collect trajectories $\{(s_t^i, a_t^i, r_t^i)\}$ by attacking the agent ${\theta_{i}}$ with a fixed adversary ${\theta'}$ in the environment $\mathcal{E}$
    
    Estimate gradient $\nabla_{\theta_i} J(\theta)$ from $\{(s_t^i, a_t^i, r_t^i)\}$
    
    Update $\theta_i$ with gradient $\nabla_{\theta_t} J(\theta_i)$
    }
\KwReturn{$\theta_N$}
\caption{Adversarial Training}\label{alg:adv}
\end{algorithm}
\begin{table*}[!htbp]
  \caption{Mean reward and the mean number of steps before blackout of the \kaist~and the \parl~across $10$ test scenarios. Each row corresponds to an adversary and each column corresponds to an agent. Standard errors across three trials are reported. Steps is defined as the episode length of safe grid operation before a blackout happens. Lower reward and steps values indicate stronger performance of the adversary.}
\makebox[\textwidth][c]{
    \begin{tabular}{ccccccc}
\toprule
          \textbf{Grid Operation RL Agent} & \multicolumn{2}{c}{\kaist~- \textit{IEEE 14-bus}} & \multicolumn{2}{c}{\kaist~- \textit{IEEE 118-bus}} & \multicolumn{2}{c}{\parl~- \textit{IEEE 118-bus}} \\
\cmidrule(lr){1-1}\cmidrule(lr){2-3}\cmidrule(lr){4-5}\cmidrule(lr){6-7}
Adversary       & Reward        & Steps         & Reward        & Steps         & Reward (thousands)        & Steps \\ \midrule
None            & $73.8 \pm 3.1$ & $813 \pm 25$ & $41.2 \pm 3.4$  & $738 \pm 26$ & $772.9 \pm 2.3$ & $864 \pm 0$ \\
Random          & $19.0 \pm 1.9$ & $343 \pm 19$ & $-67.6 \pm 0.6$ & $55 \pm 8 $& $94.7 \pm 2.6$ & $104 \pm 4$ \\
Weighted Random & $18.9 \pm 4.6$ & $322 \pm 39$ & $-69.7 \pm 1.7$ & $44 \pm 4$ & $139.9 \pm 8.2$ & $155 \pm 11$ \\
\midrule 
\textbf{Whitebox attack (ours)} & $\mathbf{-15.4 \pm 0.8}$ & $\mathbf{160 \pm 4}$ & $\mathbf{-80.8 \pm 1.8}$ & $\mathbf{35 \pm 6}$ & $\mathbf{6.0 \pm 0.8}$ & $\mathbf{7 \pm 1}$ \\
\bottomrule
\end{tabular}%
}
\label{tab:baselinecompare}%
\end{table*}

We found it extremely helpful to initialize the agent parameters $\theta$ using a pretrained agent. Thus our adversarial training procedure can be viewed as fine-tuning the parameters of the pretrained agent. We found that randomly initializing the parameters made training difficult. The agent follows a similar procedure as standard RL training, except an additional interaction with the trained adversary at each step. More specifically, we start with an L2RPN environment $\mathcal{E}$ and initial parameters of the agent $\theta_0$ and a pre-trained adversary $\theta'$. From the agent's perspective, one step in the environment is as follows: 
\begin{enumerate}
    \item Adversary observes $s_{t-1}$ and chooses action $a'_{t-1}$.
    \item Environment updates to $s_t$.
    \item Agent observes $s_t$ and chooses action $a_t$.
    \item Environment updates to $s_{t+1}$ and store $(s_t, a_t, r_t)$.
\end{enumerate} 
These rollouts are then collected and used to estimate the gradient following PPO~\cite{schulman2017ppo}. We then update $\theta_i$ in the direction of the gradient, maximizing the expected total reward.

\section{Evaluation}
We end the paper with experiments demonstrating the effectiveness of our proposed approach. 
We first describe our evaluation setup and then results from our attacks and our defense. Our code will be made publicly available which relies on the L2RPN environment~\cite{marot2020whitepaper} and the grid2op framework~\cite{grid2op}.

\subsection{Environment and Evaluation}
\textbf{Experimental Setup} We use two networks to evaluate our results on. For demonstration and visualization purposes, we use the IEEE 14 grid. 
For our main results (Tables~\ref{tab:baselinecompare},~\ref{tab:transfer}, and~\ref{tab:advcompare}) we use a subset of lines from the IEEE 118 grid~\cite{christie2000power}, directly provided in the L2RPN package~\cite{marot2020whitepaper}. These grids are much larger, with $36$ substations and $59$ power lines, resulting in around $1.88 \cdot 10^{21}$ topologies and $5.76 \cdot 10^{17}$ power line actions. Because our grids are approaching the size of real-world power grids and there is research towards scaling deep RL algorithms~\cite{stooke2018accelerated}, it is reasonable to expect that our results will hold in the real-world regime.

Each environment has its own set of test scenarios, which specify the load/generation every five minutes. Scenarios run for a maximum of 864 steps for the WCCI L2RPN challenge, or 3 days in the NeuRIPS L2RPN challenge. In each environment, we allow only \emph{a subset of the lines (around $1/6$) to be attacked}, reflecting practical constraints on the adversary's power. We found that some of the lines cause an immediate blackout when decommissioned, and we removed these lines from the adversary action set. 
For comparison convenience, we normalized reward in the range $[-100, 100]$ (except for the NeuRIPS L2RPN environment, where we did not have the necessary data to scale scores). For reference, an agent which takes no actions under no adversarial attack would receive a score of $0$ and an agent which fully optimizes the power flow would receive a score of $100.$

\textbf{Our Approach and Baselines} To study the robustness of RL agents and demonstrate the effectiveness of proposed adversarial attacks, we use the \kaist~(the winning agent from the 2020 WCCI L2RPN challenge~\cite{yoon2021winning}), the \parl~(the winning agent from the 2020 NeurIPS L2RPN challenge~\cite{liu2020parl}), the \nanyang~(the third-place agent from the 2020 WCCI L2RPN challenge~\cite{yan2020wcci}). We also use a \dqn~as a baseline, which is provided by~\cite{grid2op}.

We evaluate our learned adversary against both the \kaist~\cite{yoon2021winning} and \parl~\cite{liu2020parl} using the provided winning policies. These agents are the strongest available, as they won their respective competitions. Each adversary is allowed to inject attacks every $k = 50$ steps (adversaries cannot immediately attack). We use the following three baselines.
\begin{enumerate}
    \item \emph{No adversary;} 
    \item \emph{Random adversary, proposed by~\cite{marot2020whitepaper}.} Whenever the adversary attacks, it randomly disconnects power line in the network uniformly;
    \item \emph{Weighted-random adversary, proposed by~\cite{omnes2021adversarial}.} Whenever the adversary attacks, it disconnects each of the power line proportional to the maximum line power flow.
\end{enumerate}
When training our adversary, we use the same state representation as the \kaist, which essentially normalizes the L2RPN state observation to standard normal distribution. A full list of hyperparameters for the agent and adversary can be found in the appendix in Table~\ref{tab:hyperparams}. 

Furthermore, to demonstrate the performance of our defense method, we implement adversarial training on the \kaist. We did not use other agents since their training code were either not available or their performance was not compared to \kaist. We compare the performance of the baseline RL agent, and the RL agent with adversarial training against four different adversaries: the learnt adversary and the three baseline adversaries.  







\subsection{Learning to attack}\label{sec:attacks}
We present results for both the white-box attack and the black-box transfer attack proprosed in Section~\ref{sec:adv_mdp}. 

\textbf{White-box attacks.} Table~\ref{tab:baselinecompare} shows our results on the proposed whitebox attack. Note that each of the three columns corresponds to a different power grid, so results should not be compared across columns. As shown in Table~\ref{tab:baselinecompare}, though the winning agents achieve high reward under the no adversary setting, their performance suffers a significant drop under attacks. Notice that a random attack can cause more than 70\% performance drop, across all test scenarios. This highlights the fragility of grid control RL algorithms.

\begin{figure}[htbp]
    \centering
\includegraphics[scale=0.89]{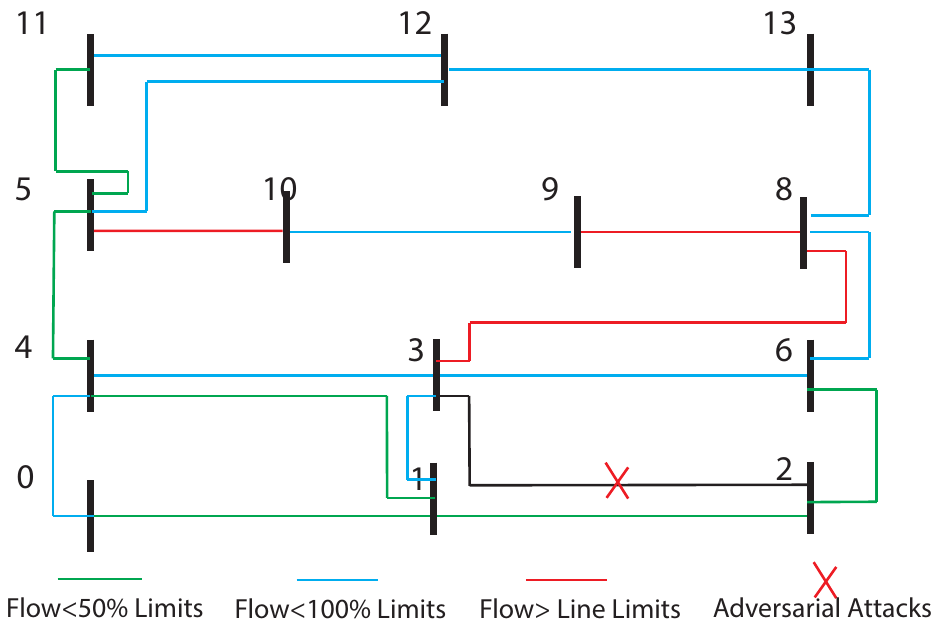}
    \caption{The learned adversary disconnects line 2-3, causing three lines (marked in red) to exceed their thermal limits. }
    \label{fig:attack}
\end{figure}

In addition, our proposed attack method is much stronger than the baseline attackers. For most runs, a trained PPO adversary is able to cause a blackout \emph{with a single attack}. In contrast, the baseline adversaries are unable to cause a blackout as effectively. 
In particular, an example attack is illustrated in Figure~\ref{fig:attack}. The learned adversary is able to disconnect the critical line (highlighted by the red cross), which causing three lines to exceed their thermal limits. A random adversary, on the other hand, typically causes no lines to overflow. Again, this example further illustrates that an RL agent without adversarial training is not robust to adversarial attacks on power grid operation.

\begin{table*}[htbp]
  \caption{The transferability of our learned adversaries across different agents on the same power grid. The row corresponds to the fixed agent used to train the adversary (see Figure~\ref{fig:method}). The column corresponds to the agent attacked by the adversary. We train three adversaries; for each one, we attack each of the agents three times and take the lowest score to measure robustness. Because we only have one agent, the clean performance has no error bars. Otherwise, we report the standard error.}
\makebox[\textwidth][c]{
    \begin{tabular}{ccccccccc}
\toprule
        \textbf{Black-box transfer attacks}   &  \multicolumn{2}{c}{\kaist~} & \multicolumn{2}{c}{\nanyang~} & \multicolumn{2}{c}{\dqn~} \\
\cmidrule(lr){1-1}\cmidrule(lr){2-3}\cmidrule(lr){4-5} \cmidrule(lr){6-7}
Adversary      & Reward        & Steps         & Reward        & Steps         & Reward     & Steps \\ \midrule
None & $47.9$ & $790$& $2.7$& $451$& $-80.5$& $128$ \\
Adv. trained using~\kaist       & $\mathbf{-81.7 \pm 0.9}$ & $\mathbf{18 \pm 0}$  & $\mathbf{-64.9 \pm 3.1}$ & $\mathbf{98 \pm 20}$ & $-91.2 \pm 0.5$ & $27 \pm 3$\\
Adv. trained using~\nanyang  & $-49.2 \pm 3.2$ & $118 \pm 6$ & $-50.6 \pm 1.1$ & $186 \pm 17$ & $-92.2 \pm 1.1$ & $27 \pm 5$ \\
Adv. trained using~\dqn        & $-23.1 \pm 0$ & $444 \pm 0$ & $-6.8 \pm 0.2$ & $409 \pm 2$ &  $\mathbf{-94.1 \pm 0}$ & $\mathbf{25 \pm 0}$ \\
\bottomrule
\end{tabular}
}
\label{tab:transfer}%
\end{table*}%

\begin{figure*}[htbp]
\centering
\centering
\includegraphics[scale=0.8]{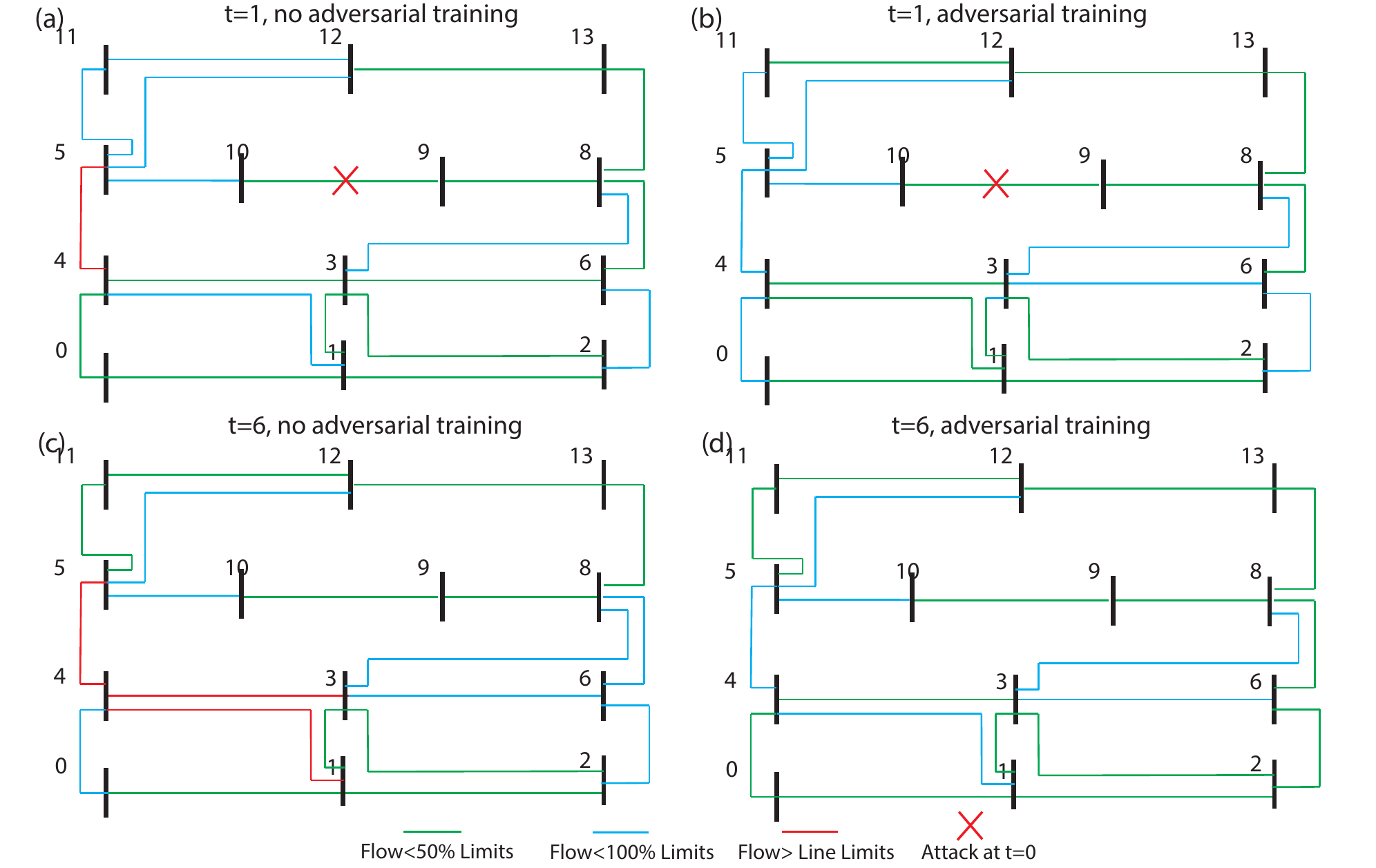}
\caption{We compare the \kaist~with and without adversarial training. At $t=0$, the line between stations $9$ and $10$ was cut in both agents. At $t=1$, the model with adversarial training was able to better distribute the load, in contrast to the model without adversarial training, where the line between $4$ and $5$ is overflowing. The overflowing line causes the model without adversarial training to destabilize at $t=6$, and afterwards it soon experiences a blackout. On the other hand, the model with adversarial training continues operating the network.}
\label{fig:defense}
\end{figure*}

\begin{table*}[htbp]
  \caption{A comparison of adversarial training using different adversaries. The row indicates the adversary used to adversarially train the agent, as described in Algorithm~\ref{alg:adv}. The column indicates the adversary used to attack the agent. For each trained model, we evaluate three times and take the lowest score to measure robustness. We also report the standard error across three trained models.}
\makebox[\textwidth][c]{
    \begin{tabular}{ccccccccc}
\toprule
\textbf{Defending attacks}& \multicolumn{2}{c}{No Adv.} & \multicolumn{2}{c}{Random Adv.} & \multicolumn{2}{c}{Weighted Random Adv.} & \multicolumn{2}{c}{Learned Adv.} \\
\cmidrule(lr){1-1}\cmidrule(lr){2-3}\cmidrule(lr){4-5}\cmidrule(lr){6-7}\cmidrule(lr){8-9}
Adv. used in adversarial training       & Reward        & Steps         & Reward        & Steps         & Reward     & Steps  & Reward     & Steps\\ \midrule
None            & $41.2 \pm 3.4$ & $738 \pm 26$ & $-72.1 \pm 1.2$ & $65 \pm 13$ & $-72.9 \pm 5.1$ & $44 \pm 4$ & $-76.5 \pm 2.6$ &  $46 \pm 10$\\
Random         & $42.2 \pm 6.9$ & $746 \pm 59$ & $-60.2 \pm 7.8$ & $102 \pm 32$ & $-51.4 \pm 0.3$ & $141 \pm 7$ & $-85.7 \pm 0.8$& $24 \pm 8$\\
Weighted Random & $44.9 \pm 1.1$ & $776 \pm 9$     & $-69.3 \pm 7.6$ & $91 \pm 40$ & $-54.9 \pm 5.3$ & $108 \pm 22$ & $-82.8 \pm 4.5$ & $39 \pm 16$\\
Learned & $\mathbf{56.3 \pm 0.1}$ & $\mathbf{864 \pm 0}$     & $\mathbf{-24.1 \pm 3.6}$ & $\mathbf{308 \pm 32}$ & $\mathbf{-16.3 \pm 1.7}$ & $\mathbf{419 \pm 6.7}$  &  $\mathbf{-39.0 \pm 4.9}$ & $\mathbf{333 \pm 16}$ \\
\bottomrule
\end{tabular}%
}
\label{tab:advcompare}%
\end{table*}%

\textbf{Black-box transfer attacks.} Table~\ref{tab:transfer} shows our results on black-box transfer attacks. First of all, as shown in Table~\ref{tab:transfer}, training against a stronger agent produces a stronger adversary. Moreover, \emph{stronger adversaries are able to transfer their attacking ability}. 
As evidenced by the second row, the adversary trained against the~\kaist produce nearly the highest performance across all agents. The high transferability highlights that the adversary is not merely exploiting agent pathologies. Instead, the learnt black-box adversary learns some concept of ``critical lines" to attack thus can lead to consistent strong attack performance.
In addition, because adversaries are able to transfer their attacking ability across agents, it implies that obfuscating the code or protecting the training algorithm of the grid operation RL agent is not a sufficient defense method for power system operators. 

We also want to point out that the \kaist~performs worse than the \nanyang~when faced with the adversary trained using both the \kaist~and the \dqn \ (see row 2 and 4 in Table~\ref{tab:transfer}). This indicates that the off-the-shelf \kaist~is likely more brittle under adversarial attacks, though it achieves higher reward in clean environment (no adversary). This observation is important in the sense that, it highlights an RL agent which appears to have \emph{stronger performance} in the absence of an adversary may actually be \emph{less robust} to potential adversary attacks. Therefore, the grid operator must include robustness into consideration, and it does not come for free with standard RL training objective. As a side note, one potential reason that the \nanyang \ achieves better robustness is that it is actually composed of two agents, and its actions is chosen from one of them under different states to maximize the expected reward. Indeed, there has been evidence showing that ensemble improves performances in other domains~\cite{pang2019ensemblerobustness, kettunen2019lpips}. It would be an interesting future direction to study how ensemble methods can help enhance RL robustness for power system operations. 

\subsection{Learning to defend}
\label{sec:defense}
Finally, we demonstrate how adversarial training can help improve the robustness of grid operation RL algorithms. 
We train using the the hyperparameters provided in~\cite{yoon2021winning}. Specially, to stabilize the adversarial training process, we first train the \kaist~for $100$ epochs without the adversary agent. Afterwards, we perform adversarial training outlined in Algorithm~\ref{alg:adv} for an additional $100$ epochs. 

Table~\ref{tab:advcompare} shows our results. We find that a stronger adversary, in turn, help find a more robust grid operation RL agent, so our learned adversary from Section~\ref{sec:attacks} is particularly useful. As a benefit of adversarial training, the clean performance (no adversary) of the \kaist~also increases. It did not suffer a blackout on any of the $10$ evaluation episodes, whereas all the other models experienced at least one. Furthermore, adversarial training allows the agent to save energy costs, as evidenced by its higher mean reward across all episodes.  

Adversarially training exposes the RL agent to more difficult scenarios than normally seen. As a result, when being attacked, it is able to quickly respond and re-balance the grid. In contrast, the agent without adversarial training struggles to do so. Eventually, these failures compound and cause more frequent blackouts. As a demonstration example, we compared how a standard \kaist~and an adversarially trained \kaist~behaved differently under same attack in Figure~\ref{fig:defense}. When an adversary disconnect a critical line, the effect quickly propagates to multiple lines after 6 time steps, which soon leads to a blackout. In contrast, the RL agent with adversarial training is able to maintain safe grid operation.

\section{Conclusion}
In this work, we look into the robustness issues of learning-based controllers in power system operation tasks. We first demonstrate an adversarial RL policy can generate  strong attacks which bring a series of operational threats to the power grid control task. By further using adversarial training to train the RL controller, we show possible routes for realizing safe RL controllers for safety-critical power grid operations.

Furthermore, we provide a realistic use-case of adversarial training, which suggests that this defense technique has the potential to be leveraged in real-world applications. We hope to encourage future work on robustness for power networks, a crucial challenge given the growing demand for electricity.

\bibliographystyle{unsrt}
\bibliography{main_pscc}

\appendix 
We used Pytorch and Tensorflow to build the models and ran the training on a NVIDIA Tesla V100 GPU with 32GB VRAM and  2.7 GHz with Dual Intel Xeon Platinum 8174 processor. 

Table~\ref{tab:hyperparams} shows the hyperparameters used for PPO~\cite{schulman2017ppo}, which was used to train the adversary. The \kaist~was trained using the hyperparameters in\cite{yoon2021winning}.
\begin{table}[htbp]
    \centering
    \begin{tabular}{c c}
        \toprule 
         Hyperparameter & Value \\
         \midrule 
         Actor and critic & Three $128$-FC layers with ReLU \\ 
         Batch size & $1024$\\ 
         Rollout length & $864$ \\
         $\gamma$ & $0.99$ \\
         Learning rate & $0.0001$ \\ 
         Clipping Range & $[0.98, 1.02]$ \\ 
         Attack frequency $k$ & $50$ steps \\
         \bottomrule
    \end{tabular}
    \caption{PPO Hyperparameters}
    \label{tab:hyperparams}
\end{table}

\end{document}